\documentclass[aps,prd,reprint,amsmath,amssymb,nofootinbib,superscriptaddress,notitlepage]{revtex4-2}

\usepackage{amsmath,amssymb,bm}
\usepackage{mathrsfs}
\usepackage[dvipsnames]{xcolor}
\usepackage{hyperref}
\hypersetup{colorlinks=true, linkcolor=Blue, citecolor=Blue, filecolor=Blue, urlcolor=Blue}

\newcommand{\Lie}{\mathscr{L}}
\newcommand{\eom}{\bumpeq}
\newcommand{\fp}{\mathrm{FP}}
\newcommand{\dd}{\mathrm{d}}

\begin{document}

\title{Bessel-Hagen currents for the Fierz-Pauli action}

\author{Michael \surname{Hobson}}
\email{mph@mrao.cam.ac.uk}
\affiliation{Astrophysics Group, Cavendish Laboratory, J.J. Thomson
  Avenue, Cambridge, CB3 0HE, UK}

\author{Will Barker}
\email{barker@fzu.cz}
\affiliation{Central European Institute for Cosmology and Fundamental Physics, Institute of Physics of the Czech Academy of Sciences, Na Slovance 1999/2, 182 00 Prague 8, Czechia}

\author{Anthony \surname{Lasenby}}
\email{a.n.lasenby@mrao.cam.ac.uk}
\affiliation{Astrophysics Group, Cavendish Laboratory, J.J. Thomson Avenue, Cambridge, CB3 0HE, UK}
\affiliation{Kavli Institute for Cosmology, Madingley Road, Cambridge, CB3 0HA, UK}

\date{\today}

\begin{abstract}
For electromagnetism in Minkowski spacetime, the Bessel-Hagen method
gives a particularly direct Noetherian derivation of the standard
gauge-invariant energy-momentum tensor. The key step is to supplement
the form variation generated by an infinitesimal coordinate
transformation with a compensating electromagnetic gauge
transformation. In this paper we ask whether the same idea can be
applied to the massless spin-2 field described by the Fierz-Pauli
action. We first prove that no nonzero local tensor quadratic in first
derivatives of the symmetric field $h_{\mu\nu}$ can be strictly
invariant under the spin-2 gauge transformation $h_{\mu\nu}\mapsto
h_{\mu\nu}+\partial_\mu\xi_\nu+\partial_\nu\xi_\mu$; the direct
electromagnetic analogue of the Bessel-Hagen construction therefore
cannot exist. Once the inexact nature of the Fierz-Pauli gauge
symmetry is treated correctly, however, the Bessel-Hagen construction
does produce a gauge-invariant equivalence class of Noether
currents. Changing the compensating spin-2 gauge parameter changes the
current only by terms proportional to the Fierz-Pauli field equations;
performing an independent spin-2 gauge transformation on $h_{\mu\nu}$
changes the current only by a trivial current given by the divergence
of an antisymmetric superpotential plus field-equation terms. This
provides the natural spin-2 analogue of Bessel-Hagen's electromagnetic
construction, but only in the quotient space of conserved currents,
and not as a preferred local gauge-invariant energy-momentum tensor.
\end{abstract}

\maketitle

\section{Introduction}

It is often stated that the Fierz-Pauli
action~\cite{FierzPauli1939} is the unique special relativistic action
for a free massless spin-2 field $h_{\mu\nu}=h_{\nu\mu}$. More precisely, it is unique up to an overall
constant and a total divergence among Lorentz-invariant Lagrangian
densities quadratic in first derivatives of $h_{\mu\nu}$ whose action
is invariant under the gauge transformation
\begin{equation}
 h_{\mu\nu}\to h'_{\mu\nu}
 =h_{\mu\nu}+\partial_\mu\xi_\nu+\partial_\nu\xi_\mu .
\label{gauge-transform}
\end{equation}
Here $\xi_\mu(x)$ is an arbitrary covector gauge parameter on Minkowski
spacetime; in the present discussion it is not to be interpreted as the
generator of a coordinate transformation.

The question considered here is whether the Bessel-Hagen
method~\cite{BesselHagen1921}, which gives the physical electromagnetic
energy-momentum tensor directly from Noether's first theorem, can similarly be
used to obtain a gauge-invariant energy-momentum tensor for the Fierz-Pauli
field. Since two different vector fields occur in the discussion, we shall
write infinitesimal coordinate transformations as
\begin{equation}
 x'^\mu=x^\mu+\zeta^\mu(x),
\label{coord-transform}
\end{equation}
whereas the spin-2 gauge transformation will be parametrised by $\xi_\mu(x)$.
We restrict attention to global Poincar\'e transformations, for which
\begin{equation}
 \zeta^\mu(x)=a^\mu+{\omega^\mu}_\nu x^\nu,
 \qquad
 \omega_{\mu\nu}=-\omega_{\nu\mu}.
\label{poincare-zeta}
\end{equation}
We use the metric signature $(+---)$ throughout.

The result obtained below has two parts. First, there is no local,
pointwise gauge-invariant tensor $\tau_{\mu\nu}\sim\partial h\,\partial h$
for the Fierz-Pauli field. The closest spin-2 precedent for this
statement is Deser and Henneaux~\cite{DeserHenneaux1995}, who showed that
the Abelian spin-2 stress tensor is conserved on the linearised Einstein
shell but is neither gauge invariant nor locally improvable to a
gauge-invariant tensor at the usual derivative order, while the
integrated Poincar\'e charges remain gauge invariant. They also
emphasise that genuine local gauge-invariant quantities must be built
from the linearised Riemann tensor, for instance the linearised
Bel-Robinson tensor. Magnano and Sokolowski~\cite{MagnanoSokolowski2002} later established
a closely related no-go theorem for the metric, or variational,
energy-momentum tensor of a gauge-invariant spin-2 field, on which
Padmanabhan relies in his review~\cite{Padmanabhan2008}; see also the
textbook discussions in Refs.~\cite{MTW1973,Wald1984}. We give a short and
explicit proof in Sec.~\ref{sec:nogo} via the observation that
$\partial h$ at a point can be set to zero by a spin-2 gauge
transformation, which fixes the geometric content of the obstruction
and connects directly to the Bessel-Hagen calculation.

Second, once the inexact gauge symmetry of the Fierz-Pauli density is
treated correctly, the Bessel-Hagen method nevertheless produces a
well-defined gauge-invariant equivalence class of Noether currents.
The equivalence relation employed here, namely currents modulo divergences
of antisymmetric superpotentials and terms proportional to the field
equations, is the standard one in the characteristic local
cohomology of Barnich, Brandt and
Henneaux (BBH)~\cite{BarnichBrandtHenneaux1995a,BarnichBrandtHenneaux1995b,BarnichBrandtHenneaux2000}.
At linearised order the non-uniqueness of the
canonical,\footnote{For the usual $\Gamma\Gamma$ representative of the
quadratic Einstein/Fierz-Pauli action, the canonical Noether tensor
coincides with the
Einstein pseudotensor at linearised order; see e.g.\ Taylor and
Baker~\cite{TaylorBaker2024} just below their Eq.~(5).}
Hilbert~\cite{Hilbert1915}, Belinfante~\cite{Belinfante1940,Rosenfeld1940}
and Landau-Lifshitz~\cite{LandauLifshitz} expressions modulo such
terms has been examined in detail by
Baker~\cite{Baker2021,BakerKK2021,BakerLinnemannSmeenk2022,LinnemannSmeenkBaker2024},
who show that infinitely many ``improved'' expressions can be
obtained from the canonical tensor by appending superpotential and
on-shell terms; a complementary parametric classification of
Noetherian energy-momentum tensors for linearised gravity has been
given by Taylor and Baker~\cite{TaylorBaker2024}.

The new content of the present paper is the
identity\footnote{The use of $\zeta$ and $\xi$ (and, later, $\chi$ and $\eta$) as
subscripts or superscripts denotes functional dependence on these
parameters and should not be confused with tensor indices.}
$J^\mu_{\zeta,\xi}=J^\mu_{\zeta,0}-2{\cal E}^{\mu\nu}\xi_\nu$, where
${\cal E}^{\mu\nu}$ is the Euler-Lagrange expression of the
Fierz-Pauli action, i.e.\ the linearised Einstein tensor up to overall
normalisation. This identity is obtained from the inexact Bessel-Hagen boundary term
$K^\mu_\xi$, together with its interpretation as a gauge-invariant
current class rather than a preferred local tensor. To the best of our
knowledge, the literature does not contain the
corresponding \emph{inexact} Bessel-Hagen calculation for the
Fierz-Pauli action, in which a single compensating spin-2 gauge
parameter $\xi_\mu$ is permitted in the form variation and the
non-vanishing gauge boundary term $K^\mu_\xi$ is included consistently
in the Noether construction. Two features of this calculation are worth flagging. (i)
The $\xi$-dependent piece $-2{\cal E}^{\mu\nu}\xi_\nu$
has precisely the weakly vanishing form familiar from the
``improvement'' literature, but
here it has a manifestly Noetherian origin; it comes from the
gauge boundary term $K^\mu_\xi$ of the inexact gauge symmetry, not
from an ad hoc improvement appended to the canonical tensor. (ii)
Within this construction, the compensating gauge
parameter $\xi_\mu$ cannot select a preferred gauge-invariant representative: it merely interchanges representatives of the same
current class.

Baker, Linnemann and Smeenk~\cite{BakerLinnemannSmeenk2022}, in Sec.~5
of their paper, identify the application of the inexact-symmetry
Bessel-Hagen method to Fierz-Pauli as the subject of future work;
Taylor and Baker~\cite{TaylorBaker2024}, in their Sec.~4.7, likewise
explicitly leave the treatment of non-exact gauge symmetries by
Bessel-Hagen's method to future work. The present
paper supplies that calculation. A close analogue had
previously been carried out by Baker and
Kuzmin~\cite{BakerKuzmin2019} for the exactly gauge-invariant case of
linearised Gauss-Bonnet gravity, where the higher-derivative,
curvature-based structure of the action allows the Bessel-Hagen
procedure to yield a strictly gauge-invariant energy-momentum tensor
even though the corresponding spin-2 form variation is not itself
gauge invariant. The Fierz-Pauli case differs essentially because the
action and current are built at the first-derivative level, where
$\partial h$ is connection-like rather than curvature-like.

\section{Electromagnetic case}
\label{sec:em}

For electromagnetism in Minkowski spacetime, the action is
\begin{equation}
 S=\int {\cal L}_{\rm EM}\,\dd^4x,
 \qquad
 {\cal L}_{\rm EM}=-\tfrac14 F_{\mu\nu}F^{\mu\nu},
\end{equation}
where $F_{\mu\nu}=\partial_\mu A_\nu-\partial_\nu A_\mu$.
Under an infinitesimal coordinate transformation $x'^\mu=x^\mu+\zeta^\mu(x)$,
the form variation of the potential is
\begin{equation}
 \delta^{(\zeta)}_0\!A_\mu
 =
 -\zeta^\nu\partial_\nu A_\mu
 -A_\nu\partial_\mu\zeta^\nu ,
\label{em-form-variation}
\end{equation}
which leaves the action invariant when $\zeta^\mu$ is given by
(\ref{poincare-zeta}). The corresponding Noether current is
\begin{equation}
 J^\mu
 =
 \frac{\partial {\cal L}_{\rm EM}}
 {\partial(\partial_\mu A_\sigma)}
 \delta^{(\zeta)}_0\!A_\sigma
 +\zeta^\mu {\cal L}_{\rm EM}.
\label{em-canonical-current}
\end{equation}
For translations
$\zeta^\mu=a^\mu$, one has  $J^\mu=-a^\nu t^\mu{}_\nu$, where
\begin{equation}
 t^\mu{}_\nu
 =
 -F^{\mu\sigma}\partial_\nu A_\sigma
 +\tfrac14\delta^\mu_\nu F^{\rho\sigma}F_{\rho\sigma}
\label{em-canonical-t}
\end{equation}
is the canonical energy-momentum tensor, which is not 
invariant under the gauge transformation
\begin{equation}
 A_\mu\to A_\mu+\partial_\mu\alpha.
\label{em-gauge-trans}
\end{equation}

The Bessel-Hagen observation is that the form variation appearing in
Noether's first theorem may be replaced by any variation that leaves
the action invariant, possibly up to a surface term. Since the electromagnetic
action is also invariant under (\ref{em-gauge-trans}), one may instead use
\begin{equation}
 \delta_0 A_\mu
 =
 -\zeta^\nu\partial_\nu A_\mu
 -A_\nu\partial_\mu\zeta^\nu
 +\partial_\mu\alpha .
\label{em-bh-variation}
\end{equation}
Choosing
\begin{equation}
 \alpha=A_\nu\zeta^\nu
\label{em-bh-choice}
\end{equation}
gives the manifestly gauge-invariant form variation
\begin{equation}
 \delta_0 A_\mu=\zeta^\nu F_{\mu\nu}.
\label{em-invariant-form-variation}
\end{equation}
The Noether current then becomes
\begin{equation}
 J^\mu
 =
 \zeta^\nu
 \left(
 F^{\mu\sigma}F_{\nu\sigma}
 -\tfrac14\delta^\mu_\nu F^{\rho\sigma}F_{\rho\sigma}
 \right)
 =
 -\zeta^\nu {\tau^\mu}_\nu ,
\end{equation}
where
\begin{equation}
 {\tau^\mu}_\nu
 =
 -\left(
 F^{\mu\sigma}F_{\nu\sigma}
 -\tfrac14\delta^\mu_\nu F^{\rho\sigma}F_{\rho\sigma}
 \right)
\end{equation}
is the standard physical energy-momentum tensor of the electromagnetic
field (the sign reflecting our metric convention). The important point is
that the compensating gauge transformation makes the form variation itself
gauge invariant. Since the other quantities appearing in the electromagnetic
current are also gauge invariant, the resulting current is gauge invariant
as a local expression.

Changing $\alpha$ in \eqref{em-bh-variation} also generates an
equivalence class of conserved currents. Writing
$J^\mu_{\zeta,\alpha}$ for the Noether current associated with a
given $\alpha$, and using the exact gauge invariance of the
electromagnetic action,
\begin{equation}
 J^\mu_{\zeta,\alpha} - J^\mu_{\zeta,0}
 =
 -F^{\mu\nu}\partial_\nu\alpha
 =
 -\partial_\nu\!\left(F^{\mu\nu}\alpha\right)
 +\left(\partial_\nu F^{\mu\nu}\right)\alpha ,
\label{em-shift}
\end{equation}
which is the divergence of an antisymmetric superpotential
$F^{\mu\nu}\alpha$ plus a term proportional to the Maxwell equations
of motion. Changing $\alpha$ therefore shifts the current only within
an equivalence class of Noether currents. What is special about the
electromagnetic case is that the natural choice \eqref{em-bh-choice}
produces a unique representative that is strictly gauge invariant as a local
tensor. As we shall see, the
Fierz-Pauli analogue treated below also admits an equivalence class of
Noether currents but no such gauge-invariant local representative.

\section{Fierz-Pauli action}

For the massless spin-2 field we take the Fierz-Pauli density in
the form\footnote{The form of ${\cal L}_{\fp}$ given in
\eqref{fp-lagrangian} is unique modulo total divergences and the
one-parameter family of linear field redefinitions
$h_{\mu\nu}\to h_{\mu\nu}+\lambda\eta_{\mu\nu}h$ that mix the trace
into the tensor. Such a redefinition rescales the trace as
$h\to(1+4\lambda)h$ in four dimensions and correspondingly alters the
coefficients of the second-order (kinetic) operator $O$ defined by
${\cal E}^{\alpha\beta}=O^{\alpha\beta\gamma\delta}h_{\gamma\delta}$
[cf.\ \eqref{E-explicit}], while leaving the massless spin-2 content
unchanged. It is invertible except at the trace-removing value
$\lambda=-\tfrac14$, where $h\to0$ and the map degenerates; this value
is therefore excluded.} (the Singh-Hagen tuned mass
term~\cite{SinghHagen1974}, which gives the massive theory at
the cost of breaking gauge invariance, is omitted throughout):
\begin{equation}
\begin{split}
 {\cal L}_{\fp}
 = \tfrac14 \big(
 & -\partial_\lambda h\,\partial^\lambda h
 +\partial_\lambda h_{\alpha\beta}
  \partial^\lambda h^{\alpha\beta}                                 \\
 & -2\partial_\lambda h_{\alpha\beta}
  \partial^\beta h^{\alpha\lambda}
 +2\partial_\alpha h\,\partial_\beta h^{\beta\alpha}
 \big),
\end{split}
\label{fp-lagrangian}
\end{equation}
where $h=h^\mu{}_\mu$. Its variation may be written as
\begin{equation}
 \delta {\cal L}_{\fp}
 =
 {\cal E}^{\alpha\beta}\delta h_{\alpha\beta}
 +\partial_\mu\theta^\mu(\delta h),
\label{fp-variation}
\end{equation}
where ${\cal E}^{\alpha\beta}\equiv\delta{\cal L}_{\fp}/\delta h_{\alpha\beta}$
is the Euler-Lagrange expression\footnote{Strictly, the variational-derivative notation
$\delta/\delta h_{\alpha\beta}$ applies to functionals rather
than to the local density ${\cal L}_{\fp}$; here it is used as
a shorthand for the Euler-Lagrange expression $\partial
{\cal L}_{\fp}/\partial h_{\alpha\beta}-\partial_\mu(\partial
{\cal L}_{\fp}/\partial(\partial_\mu h_{\alpha\beta}))$.} of the Fierz-Pauli action, given
explicitly in \eqref{E-explicit} below (it is the linearised Einstein
tensor, whose
vanishing gives the linearised Einstein equation for $h_{\alpha\beta}$), and
\begin{equation}
 \theta^\mu(\delta h)
 =
 \Pi^{\mu|\alpha\beta}\delta h_{\alpha\beta},
 \qquad
 \Pi^{\mu|\alpha\beta}
 \equiv
 \frac{\partial{\cal L}_{\fp}}
 {\partial(\partial_\mu h_{\alpha\beta})}.
\label{symplectic-potential}
\end{equation}
Direct differentiation of \eqref{fp-lagrangian} gives
\begin{equation}
\begin{split}
 \Pi^{\mu|\alpha\beta}
 =&\;
 \tfrac12\partial^\mu h^{\alpha\beta}
 -\tfrac12\eta^{\alpha\beta}\partial^\mu h
 -\partial^{(\alpha}h^{\beta)\mu}       \\
 &\;
 +\tfrac12\eta^{\alpha\beta}\partial_\lambda h^{\lambda\mu}
 +\tfrac12\eta^{\mu(\alpha}\partial^{\beta)}h ,
\end{split}
\label{canonical-momentum}
\end{equation}
where parentheses around indices denote symmetrisation with unit
weight, $A^{(\alpha\beta)}=\tfrac12(A^{\alpha\beta}+A^{\beta\alpha})$.
Since ${\cal L}_{\fp}$ has no explicit dependence on
$h_{\alpha\beta}$, the Euler-Lagrange expression is
\begin{equation}
\begin{split}
 {\cal E}^{\mu\nu}
 =
 -\partial_\sigma\Pi^{\sigma|\mu\nu}
 =\;&
 -\tfrac12\Box h^{\mu\nu}
 +\partial_\sigma\partial^{(\mu}h^{\nu)\sigma}
 -\tfrac12\partial^\mu\partial^\nu h \\
 &
 +\tfrac12\eta^{\mu\nu}\!\left(
 \Box h - \partial_\sigma\partial_\rho h^{\rho\sigma}
 \right),
\end{split}
\label{E-explicit}
\end{equation}
which is the linearised Einstein tensor.
The
gauge invariance of the action under
\begin{equation}
 D_\xi h_{\alpha\beta}
 \equiv
 \partial_\alpha\xi_\beta+\partial_\beta\xi_\alpha
\label{Dxi}
\end{equation}
implies the Noether identity
\begin{equation}
 \partial_\alpha {\cal E}^{\alpha\beta}\equiv0 .
\label{noether-identity}
\end{equation}

Under a Poincar\'e coordinate transformation $x'^\mu=x^\mu+\zeta^\mu$, the
form variation of $h_{\alpha\beta}$ is
\begin{equation}
 \delta^{(\zeta)}_0\!h_{\alpha\beta}
 =
 -\zeta^\rho\partial_\rho h_{\alpha\beta}
 -h_{\rho\beta}\partial_\alpha\zeta^\rho
 -h_{\alpha\rho}\partial_\beta\zeta^\rho .
\label{fp-form-variation}
\end{equation}
As in the electromagnetic case, the Bessel-Hagen observation is that
the form variation appearing in Noether's first theorem may be
replaced by any variation that leaves the action invariant. Since the
Fierz-Pauli action is also invariant under the spin-2 gauge
transformation \eqref{gauge-transform}, one may augment
\eqref{fp-form-variation} by the compensating piece $D_\xi h_{\alpha\beta}
=\partial_\alpha\xi_\beta+\partial_\beta\xi_\alpha$ and use
\begin{equation}
\begin{split}
 \delta_0 h_{\alpha\beta}
 =\;& -\zeta^\rho\partial_\rho h_{\alpha\beta}
       -h_{\rho\beta}\partial_\alpha\zeta^\rho
       -h_{\alpha\rho}\partial_\beta\zeta^\rho \\
     & +\partial_\alpha\xi_\beta+\partial_\beta\xi_\alpha .
\end{split}
\label{fp-bh-variation}
\end{equation}
We shall write this combined form variation compactly as
\begin{equation}
 \Delta^{\zeta,\xi}_{\alpha\beta}
 \equiv
 -\Lie_\zeta h_{\alpha\beta}
 +D_\xi h_{\alpha\beta},
\label{Q-def}
\end{equation}
where $\Lie_\zeta h_{\alpha\beta} = \zeta^\rho\partial_\rho
h_{\alpha\beta} +h_{\rho\beta}\partial_\alpha\zeta^\rho
+h_{\alpha\rho}\partial_\beta\zeta^\rho$ is the Lie derivative of
$h_{\alpha\beta}$ regarded as a $(0,2)$ tensor on Minkowski spacetime,
so that $-\Lie_\zeta
h_{\alpha\beta}=\delta^{(\zeta)}_0\!h_{\alpha\beta}$ recovers the
Poincar\'e piece \eqref{fp-form-variation}.

The most immediate
analogue of the electromagnetic choice (\ref{em-bh-choice}) is
\begin{equation}
 \xi_\mu=h_{\mu\nu}\zeta^\nu .
\label{naive-xi}
\end{equation}
A short calculation gives
\begin{equation}
\begin{split}
 \Delta^{\zeta,h\zeta}_{\alpha\beta}
 &=
 \zeta^\rho
 \left(
 \partial_\alpha h_{\beta\rho}
 +\partial_\beta h_{\alpha\rho}
 -\partial_\rho h_{\alpha\beta}
 \right)                                      \\
 &=
 2\zeta^\rho\,\Gamma^{(1)}_{\rho\alpha\beta},
\end{split}
\label{Q-connection}
\end{equation}
where
$\Gamma^{(1)}_{\rho\alpha\beta}\equiv\tfrac12(\partial_\alpha h_{\beta\rho}+\partial_\beta h_{\alpha\rho}-\partial_\rho h_{\alpha\beta})$
is the linearised Christoffel symbol. This is the closest spin-2
analogue of the electromagnetic result; it coincides with the
linearised Christoffel form variation studied by Baker and
Kuzmin~\cite{BakerKuzmin2019} in the linearised Gauss-Bonnet context
and isolated in Sec.~4.3 of Taylor and Baker~\cite{TaylorBaker2024}
within their general parametric classification of form variations for
linearised gravity. There is, however, a crucial difference between EM and
Fierz-Pauli: $F_{\mu\nu}$ is gauge invariant, whereas
$\Gamma^{(1)}_{\rho\alpha\beta}$ is not. Indeed, under an independent
spin-2 gauge transformation
$h_{\mu\nu}\to h_{\mu\nu}+D_\chi h_{\mu\nu}$ with parameter $\chi_\mu$
(distinct from the compensating $\xi_\mu$ just chosen),
\begin{equation}
 \Gamma^{(1)}_{\rho\alpha\beta}
 \to
 \Gamma^{(1)}_{\rho\alpha\beta}
 +\partial_\alpha\partial_\beta\chi_\rho .
\label{connection-gauge}
\end{equation}

\section{Local no-go result}
\label{sec:nogo}

The lack of gauge invariance just described is not merely the result
of a poor choice (\ref{naive-xi}) of compensating gauge parameter. No local tensor
quadratic in first derivatives of $h_{\mu\nu}$ can be strictly gauge
invariant. The argument is the linearised gauge-theoretic analogue
of the familiar fact that all Christoffel symbols vanish at a point
in Riemann normal coordinates.

To see this, work at a point $p$. The first derivative
$\partial_\lambda h_{\mu\nu}$ at $p$ is equivalent to specifying the
linearised connection $\Gamma^{(1)}_{\rho\mu\nu}$ at $p$, since
\begin{equation}
 \partial_\lambda h_{\mu\nu}
 =
 \Gamma^{(1)}_{\mu\lambda\nu}
 +\Gamma^{(1)}_{\nu\lambda\mu}.
\label{H-Gamma-inverse}
\end{equation} Under the spin-2 gauge transformation,
\begin{equation}
 \delta\Gamma^{(1)}_{\rho\mu\nu}
 =
 \partial_\mu\partial_\nu\xi_\rho .
\end{equation}
At the point $p$ the symmetric second derivatives
$\partial_\mu\partial_\nu\xi_\rho(p)$ are arbitrary independent quantities.
Hence, given any value of $\Gamma^{(1)}_{\rho\mu\nu}(p)$, one may choose
$\xi_\rho$ so that
\begin{equation}
 \Gamma^{(1)}_{\rho\mu\nu}(p)\to 0,
\end{equation}
and therefore also $\partial_\lambda h_{\mu\nu}(p)\to 0$.

Suppose now that $\tau_{\mu\nu}$ is a local pointwise expression
depending only on $\eta_{\mu\nu}$ and $\partial_\lambda h_{\mu\nu}$,
homogeneous quadratic in $\partial h$, transforming as a Lorentz
tensor, and strictly invariant under the spin-2 gauge transformation. Since every value of $\partial h$ at $p$ is
gauge-equivalent to zero,
\begin{equation}
 \tau_{\mu\nu}(\partial h)=\tau_{\mu\nu}(0).
\end{equation}
But $\tau_{\mu\nu}$ is quadratic in $\partial h$, so
$\tau_{\mu\nu}(0)=0$. Therefore
\begin{equation}
 \tau_{\mu\nu}\equiv 0 .
\end{equation}
This proves the claim: there is no nonzero local, pointwise gauge-invariant
tensor $\tau_{\mu\nu}\sim\partial h\,\partial h$ for the Fierz-Pauli field.
The first genuinely gauge-invariant local field strength is the linearised
Riemann tensor,
\begin{equation}
 R^{(1)}_{\mu\nu\rho\sigma}\sim \partial\partial h ,
\end{equation}
so strictly gauge-invariant local quantities can be built from
$R^{(1)}_{\mu\nu\rho\sigma}$ and its derivatives, for instance the
linearised Bel-Robinson tensor quadratic in $R^{(1)}$, but not from
$\partial h\,\partial h$ alone~\cite{DeserHenneaux1995}.

\section{Inexact Bessel-Hagen current}
\label{sec:inexact}

Although the preceding result precludes a strict electromagnetic
analogue, it does not rule out a weaker Noetherian result. The
important point is that the Fierz-Pauli gauge symmetry is an inexact
symmetry of the Lagrangian density: the action is invariant, but the
density changes by a total divergence. Although this distinction is
invisible at the level of the action, it is essential at the level of
the current. Noether's construction builds the conserved current
from the density-level transformation, so the boundary term
$K^\mu_\xi$ defined below enters the current explicitly. For
electromagnetism the corresponding boundary term vanishes identically,
since $\delta_\alpha{\cal L}_{\rm EM}=0$; for Fierz-Pauli it does not,
and it is precisely this term that produces the $\xi$-dependent piece
of the current.

For a pure spin-2 gauge transformation
$\delta h_{\alpha\beta}=D_\xi h_{\alpha\beta}$, the variational identity
\eqref{fp-variation} gives
\begin{equation}
 \delta_\xi {\cal L}_{\fp}
 =
 {\cal E}^{\alpha\beta}D_\xi h_{\alpha\beta}
 +\partial_\mu\theta^\mu(D_\xi h).
\end{equation}
Using \eqref{noether-identity}, the first term may be written as
\begin{equation}
 {\cal E}^{\alpha\beta}D_\xi h_{\alpha\beta}
 =
 2{\cal E}^{\alpha\beta}\partial_\alpha\xi_\beta
 =
 \partial_\alpha\!\left(2{\cal E}^{\alpha\beta}\xi_\beta\right).
\end{equation}
Thus
\begin{equation}
 \delta_\xi {\cal L}_{\fp}=\partial_\mu K^\mu_\xi,
\end{equation}
which determines $K^\mu_\xi$ only up to the divergence of an
antisymmetric superpotential. The natural representative,
obtained by reading the divergence directly from the steps above,
is
\begin{equation}
 K^\mu_\xi
 =
 \theta^\mu(D_\xi h)
 +2{\cal E}^{\mu\nu}\xi_\nu .
\label{K-def}
\end{equation}
Any other choice differs from this by an antisymmetric-superpotential
divergence and shifts the resulting Bessel-Hagen current within its
BBH equivalence class.

The calculation below is first carried out for an arbitrary
external gauge parameter $\xi_\mu(x)$. Field-dependent Bessel-Hagen choices such as
$\xi_\mu=h_{\mu\nu}\zeta^\nu$ are then obtained by substitution
(the derivation uses only $\partial_\mu{\cal E}^{\mu\nu}\equiv 0$
and is therefore insensitive to whether $\xi_\mu$ depends on $h$);
their field-dependence is accounted for explicitly when an
independent gauge variation $\delta_\chi$ is later taken in
Sec.~\ref{sec:class}.

For a combined Poincar\'e and spin-2 gauge transformation
$\delta_0 h=\Delta^{\zeta,\xi}$ in \eqref{Q-def}, the variation of the density is
$\delta_0 {\cal L}_{\fp}=-\partial_\mu(\zeta^\mu{\cal L}_{\fp})
+\partial_\mu K^\mu_\xi$, since $\partial_\mu\zeta^\mu=0$ for Poincar\'e.
Combining with \eqref{fp-variation} gives
\begin{equation}
 \partial_\mu\!\left[
 \theta^\mu(\Delta^{\zeta,\xi})+\zeta^\mu{\cal L}_{\fp}-K^\mu_\xi
 \right]
 =
 -{\cal E}^{\alpha\beta}\Delta^{\zeta,\xi}_{\alpha\beta}.
\end{equation}
Since ${\cal E}^{\alpha\beta}\eom 0$ on the Fierz-Pauli equations of
motion --- we use $\eom$ throughout for weak equality, i.e.\
equality modulo the field equations --- the right-hand side
vanishes on-shell, identifying the bracket on the left as the
conserved Bessel-Hagen current corresponding to the combined
transformation:
\begin{equation}
 J^\mu_{\zeta,\xi}
 =
 \theta^\mu(\Delta^{\zeta,\xi})
 +\zeta^\mu{\cal L}_{\fp}
 -K^\mu_\xi .
\label{BH-current}
\end{equation}
Substituting \eqref{K-def} and using linearity of $\theta^\mu$ in its
argument gives the simple form
\begin{equation}
 J^\mu_{\zeta,\xi}
 =
 \theta^\mu(-\Lie_\zeta h)
 +\zeta^\mu{\cal L}_{\fp}
 -2{\cal E}^{\mu\nu}\xi_\nu .
\label{central-current}
\end{equation}
On-shell, this becomes
\begin{equation}
 J^\mu_{\zeta,\xi}
 \eom
 \theta^\mu(-\Lie_\zeta h)
 +\zeta^\mu{\cal L}_{\fp}.
\label{current-onshell}
\end{equation}
This is the central result. Equivalently, writing
$J^\mu_{\zeta,0}\equiv\theta^\mu(-\Lie_\zeta h)+\zeta^\mu{\cal L}_{\fp}$
for the $\xi=0$ representative, Eq.~\eqref{central-current} reads
\begin{equation}
 J^\mu_{\zeta,\xi}
 =
 J^\mu_{\zeta,0}
 -2{\cal E}^{\mu\nu}\xi_\nu .
\label{xi-shift}
\end{equation}
Note first that \eqref{xi-shift} is an off-shell equality of
local tensors, not merely an on-shell equivalence: although the
right-hand side has the same general form as the EOM-improvement
freedom of the BBH equivalence relation, the specific shift
$-2{\cal E}^{\mu\nu}\xi_\nu$ is an off-shell prediction of the
Bessel-Hagen construction, not a freedom merely permitted by that
equivalence relation. Indeed, although $K^\mu_\xi$ admits superpotential
additions, the only freedom propagating through the construction is that
of the symplectic potential $\theta^\mu$, which shifts $J^\mu_{\zeta,\xi}$
and $J^\mu_{\zeta,0}$ identically and so cancels in their difference; the
$\xi$-dependence is thus $-2{\cal E}^{\mu\nu}\xi_\nu$ independently of that
choice. This term is not itself a superpotential: by
$\partial_\mu{\cal E}^{\mu\nu}\equiv0$ its divergence is
$\partial_\mu(-2{\cal E}^{\mu\nu}\xi_\nu)=-{\cal E}^{\mu\nu}D_\xi h_{\mu\nu}$,
the (on-shell-vanishing) Euler-Lagrange contraction, whereas a
superpotential is identically conserved. The shift therefore lies in the
weakly-vanishing sector of the trivial currents rather than the
superpotential sector, which is the precise sense in which \eqref{xi-shift}
is an off-shell statement and not merely a relation within the BBH class.
Two further consequences of \eqref{xi-shift} are worth recording. First, the
identity holds for any local choice $\xi_\mu[h,\zeta]$, since its
derivation uses only $\partial_\mu{\cal E}^{\mu\nu}\equiv 0$; the
electromagnetic-like choice $\xi_\mu=h_{\mu\nu}\zeta^\nu$ is singled
out only because it makes the form variation connection-like,
$\Delta^{\zeta,h\zeta}_{\mu\nu}=2\zeta^\rho\Gamma^{(1)}_{\rho\mu\nu}$, not
at the level of the current class. Second, the $\xi$-dependent piece
$-2{\cal E}^{\mu\nu}\xi_\nu$ is \emph{not} an ad hoc improvement: it
arises from the gauge boundary term $K^\mu_\xi$ via
Eq.~\eqref{K-def} and is therefore built into the Bessel-Hagen
construction itself. The $\xi$-family parametrised by
\eqref{xi-shift} spans only the equation-of-motion half of the
equivalence class; the external superpotential plus on-shell
freedom of
Refs.~\cite{Baker2021,BakerKK2021,BakerLinnemannSmeenk2022,LinnemannSmeenkBaker2024}
enters via the BBH equivalence relation invoked in
Sec.~\ref{sec:class}.

The current \eqref{central-current} is conserved only weakly,
$\partial_\mu J^\mu_{\zeta,\xi}\eom0$, as is standard for Noether currents
and as already holds for the Maxwell tensor of Sec.~\ref{sec:em}; the
off-shell content of the construction resides in the local identity
\eqref{xi-shift} rather than in the conservation law. The physically
meaningful object is the conserved charge $Q_\zeta=\int_\Sigma
J^0_{\zeta,\xi}\,\dd^3x$, which is conserved on solutions and is
independent of the compensating parameter $\xi$, since the shift
$-2{\cal E}^{\mu\nu}\xi_\nu$ vanishes on shell. The attendant
non-uniqueness of the local current---its determination only up to the
equivalence class---is thus not a limitation peculiar to this construction
but the expected feature of gravitational energy-momentum, with the charge
as the well-defined invariant.

Relative to the parametric classification of Taylor and
Baker~\cite{TaylorBaker2024}, the point is therefore the following.
The linearised-Christoffel form variation \eqref{Q-connection} is one
member of their family of admissible form variations, but the
current-level identity \eqref{xi-shift} depends on the inexact gauge
boundary term $K^\mu_\xi$, which lies outside their exact-symmetry
classification.

Choosing the $\xi=0$ representative of the class and specialising to
translations $\zeta^\mu=a^\mu$, one obtains
\begin{equation}
 J^\mu_{a,0}
 =
 -a^\rho\Pi^{\mu|\alpha\beta}\partial_\rho h_{\alpha\beta}
 +a^\mu{\cal L}_{\fp}
 =
 -a^\rho\,{t^\mu}_\rho,
\end{equation}
where
\begin{equation}
 {t^\mu}_\rho
 =
 \Pi^{\mu|\alpha\beta}\partial_\rho h_{\alpha\beta}
 -\delta^\mu_\rho{\cal L}_{\fp}
\label{canonical-t}
\end{equation}
is the canonical energy-momentum tensor of the Fierz-Pauli field.
This tensor is not gauge invariant as a local expression, but the
current class it defines \emph{is} gauge invariant in the sense made
precise in Sec.~\ref{sec:class} below.

We have displayed the translation current because it is the part of
the Poincar\'e current conventionally identified with energy-momentum.
Nothing in the preceding derivation is restricted to translations,
however. For Lorentz transformations
$\zeta^\mu={\omega^\mu}_\nu x^\nu$, the same identity
\eqref{xi-shift} holds and gives the corresponding angular-momentum
current, schematically $J^\mu_\omega=\tfrac12\omega_{\rho\sigma}
(x^\rho t^{\mu\sigma}-x^\sigma t^{\mu\rho}+S^{\mu|\rho\sigma})$,
where $S^{\mu|\rho\sigma}$ is the spin current associated with the
tensor indices of $h_{\mu\nu}$. Since the present paper is concerned
with the energy-momentum tensor, we spell out only the translation
specialisation.

\section{Gauge invariance of the current class}
\label{sec:class}

Section~\ref{sec:inexact} showed that varying the compensating
Bessel-Hagen parameter $\xi$, with the field $h_{\mu\nu}$ held fixed,
keeps the current within a fixed equivalence class. We now show that
this class is itself gauge invariant under an independent
transformation $h_{\mu\nu}\to h_{\mu\nu}+D_\chi h_{\mu\nu}$ of the
field. The current \eqref{BH-current} satisfies the off-shell
identity
\begin{equation}
 \partial_\mu J^\mu_{\zeta,\xi}
 =
 -{\cal E}^{\alpha\beta}\Delta^{\zeta,\xi}_{\alpha\beta},
\label{current-divergence}
\end{equation}
which is just Noether's identity for the combined Poincar\'e and
spin-2 gauge variation. Now perform an independent spin-2 gauge
transformation with parameter $\chi_\mu$, distinct from the
compensating $\xi_\mu$ already appearing in $\Delta^{\zeta,\xi}$:
\begin{equation}
 \delta_\chi h_{\mu\nu}=D_\chi h_{\mu\nu}.
\end{equation}
Since the Fierz-Pauli Euler expression is gauge invariant,
$\delta_\chi{\cal E}^{\alpha\beta}=0$, we obtain from
\eqref{current-divergence}
\begin{equation}
 \partial_\mu\delta_\chi J^\mu_{\zeta,\xi}
 =
 -{\cal E}^{\alpha\beta}\delta_\chi \Delta^{\zeta,\xi}_{\alpha\beta}.
\label{gauge-var-divergence}
\end{equation}
By \eqref{gauge-var-divergence}, the induced change in the current is
controlled entirely by the change in the form variation
$\Delta^{\zeta,\xi}$, which we now evaluate. Writing
$\Delta^{\zeta,\xi}_{\mu\nu}=-\Lie_\zeta h_{\mu\nu}+D_\xi h_{\mu\nu}$, the
independent transformation $\delta_\chi$ acts through two channels:
directly on the field $h_{\mu\nu}$ in the transport term $-\Lie_\zeta h$,
and on the field-dependent compensating parameter $\xi_\mu[h,\zeta]$ in
$D_\xi h$. Together these give
\begin{equation}
 \delta_\chi \Delta^{\zeta,\xi}_{\mu\nu}
 =
 -\Lie_\zeta(D_\chi h_{\mu\nu})
 +D_{\delta_\chi\xi}h_{\mu\nu} .
\end{equation}
The first channel can be simplified: for the Poincar\'e transformations
considered here, $\partial_\mu\partial_\nu\zeta^\rho=0$, so that the Lie
derivative along $\zeta$ commutes with the spin-2 gauge action up to a
relabelling of the parameter, $\chi\to\Lie_\zeta\chi$, and hence
\begin{equation}
 \Lie_\zeta(D_\chi h_{\mu\nu})
 =
 D_{\Lie_\zeta\chi}h_{\mu\nu} ,
\label{commutation-identity}
\end{equation}
where $\Lie_\zeta\chi$ denotes the Lie derivative of $\chi_\mu$
regarded as a covector gauge parameter on Minkowski spacetime,
$(\Lie_\zeta\chi)_\mu=\zeta^\rho\partial_\rho\chi_\mu+\chi_\rho\partial_\mu\zeta^\rho$.
Defining the effective spin-2 gauge parameter
$\eta_\mu\equiv\delta_\chi\xi_\mu-(\Lie_\zeta\chi)_\mu$, which
combines the induced change of the compensating $\xi[h]$ under the
$\chi$-transformation of the field and the failure of the Poincar\'e
and gauge actions to commute on $h$, it follows that
\begin{equation}
 \delta_\chi \Delta^{\zeta,\xi}_{\mu\nu}
 =
 D_\eta h_{\mu\nu} .
\label{eta-def}
\end{equation}
Equation \eqref{eta-def} is the crucial step: the gauge variation of the
form variation is itself a pure spin-2 gauge variation. It is precisely
because $\delta_\chi\Delta^{\zeta,\xi}$ is again of pure-gauge form that the
induced change in the current is a trivial (pure-gauge) current, as we now
show. The Noether current associated with the pure spin-2 gauge
variation $D_\eta h_{\mu\nu}$ is trivial:
\begin{equation}
 J^\mu_\eta
 =
 \theta^\mu(D_\eta h)-K^\mu_\eta
 =
 -2{\cal E}^{\mu\nu}\eta_\nu .
\label{pure-gauge-current}
\end{equation}
Indeed, using the Noether identity
$\partial_\mu{\cal E}^{\mu\nu}\equiv0$, its divergence is
\begin{equation}
 \partial_\mu J^\mu_\eta
 =
 -2{\cal E}^{\mu\nu}\partial_\mu\eta_\nu
 =
 -{\cal E}^{\mu\nu}D_\eta h_{\mu\nu}.
\end{equation}
Equations \eqref{gauge-var-divergence} and \eqref{eta-def} therefore
imply
\begin{equation}
 \partial_\mu\!\left(
 \delta_\chi J^\mu_{\zeta,\xi}-J^\mu_\eta
 \right)\equiv0 .
\end{equation}
Thus, locally in a topologically trivial region, the difference is an
identically conserved current and hence is the divergence of an
antisymmetric superpotential,\footnote{This is the local algebraic
Poincar\'e lemma for identically conserved currents. Equivalently, it
is the standard trivial-current equivalence relation used in
characteristic
cohomology~\cite{BarnichBrandtHenneaux1995a,BarnichBrandtHenneaux1995b,BarnichBrandtHenneaux2000}:
currents differing by superpotentials and weakly vanishing currents
represent the same conservation law.}
\begin{equation}
 \delta_\chi J^\mu_{\zeta,\xi}-J^\mu_\eta
 =
 \partial_\nu U^{[\mu\nu]}_{\zeta,\chi} .
\end{equation}
Substituting \eqref{pure-gauge-current} gives
\begin{equation}
 \delta_\chi J^\mu_{\zeta,\xi}
 =
 \partial_\nu U^{[\mu\nu]}_{\zeta,\chi}
 -2{\cal E}^{\mu\nu}
 \big[
 \delta_\chi\xi_\nu-(\Lie_\zeta\chi)_\nu
 \big].
\label{current-gauge-variation}
\end{equation}
Consequently, on shell
\begin{equation}
 \delta_\chi J^\mu_{\zeta,\xi}
 \eom
 \partial_\nu U^{[\mu\nu]}_{\zeta,\chi}.
\label{weak-gauge-invariance}
\end{equation}
This is precisely the expected notion of gauge invariance for a Noether
current. Currents which differ by
\begin{equation}
 J^\mu
 \to
 J^\mu
 +\partial_\nu U^{[\mu\nu]}
 +V^\mu{}_{\alpha\beta}{\cal E}^{\alpha\beta}
\label{current-equivalence}
\end{equation}
define the same conserved charge under the usual boundary conditions.
We denote the resulting equivalence class by $[\cdot]$. The object
produced by the Bessel-Hagen method for the Fierz-Pauli action is
therefore not a single gauge-invariant local tensor, but the
equivalence class
\begin{equation}
 [J^\mu_\zeta]
 =
 \left[
 \theta^\mu(-\Lie_\zeta h)
 +\zeta^\mu{\cal L}_{\fp}
 \right].
\label{current-class}
\end{equation}

\section{Discussion}

The contrast with electromagnetism is now clear. In electromagnetism the
gauge-invariant field strength is first order in the potential,
$F_{\mu\nu}\sim\partial A$, and the Bessel-Hagen choice
$\alpha=A_\mu\zeta^\mu$ turns the form variation itself into the gauge-invariant
expression $\zeta^\nu F_{\mu\nu}$. The Noether current therefore 
produces a strictly gauge-invariant local representative of its equivalence
class: the standard Maxwell energy-momentum tensor.

For the Fierz-Pauli field, the first derivative $\partial h$ is
connection-like. It can be made to vanish at a point by a spin-2 gauge
transformation. The first local gauge-invariant field strength is
curvature-like and contains two derivatives of $h_{\mu\nu}$. Thus no
Bessel-Hagen choice can give a nonzero local tensor of the form
$\partial h\,\partial h$ that is strictly gauge invariant. This explains
why the numerous canonical, Belinfante, Hilbert, Landau-Lifshitz and
related expressions for gravitational energy in linearised gravity are
either gauge dependent, defined only after gauge fixing, or equivalent only
after quotienting by superpotentials and field-equation terms.

The calculation above shows, however, that the Bessel-Hagen method
does not simply fail for Fierz-Pauli theory. Rather, once the inexact
gauge symmetry of the Fierz-Pauli density is treated correctly, the
method naturally lands in the quotient space of Noether
currents.\footnote{A concrete illustration is provided by the
Butcher tensor of
Refs.~\cite{ButcherHobsonLasenby2010,ButcherLasenbyHobson2012,ButcherHobsonLasenby2012},
originally introduced in harmonic gauge in
Ref.~\cite{ButcherHobsonLasenby2010} and later generalised to
other gauges in Ref.~\cite{ButcherHobsonLasenby2012}. For its
original harmonic-gauge form, Barker
\emph{et al.}~\cite{BarkerLasenbyHobsonHandley2019} showed that
it is obtained from the linearised Einstein pseudotensor (in the
same gauge) by subtracting an identically conserved current,
i.e.\ the divergence of an antisymmetric superpotential. This is
therefore a relocalisation within the equivalence class of
Noether currents considered in the present paper.} The
compensating spin-2 gauge parameter $\xi_\mu$ cannot be used to
select a preferred gauge-invariant local representative, but it also cannot alter
the Noether current class. This provides a precise spin-2
counterpart of the Bessel-Hagen construction in electromagnetism, albeit
in a weaker sense.

It may seem at first sight that the construction has merely returned
the canonical Noether energy-momentum tensor. The point, however, is
not that \eqref{canonical-t} is a new local expression; it is not.
Rather, the inexact Bessel-Hagen construction shows
why no choice of compensating spin-2 gauge parameter can improve this
representative into a preferred gauge-invariant local tensor: all
such choices give
$J^\mu_{\zeta,\xi}=J^\mu_{\zeta,0}-2{\cal E}^{\mu\nu}\xi_\nu$, and
hence the same current class. The canonical tensor appears here only
as a convenient representative of the Bessel-Hagen current class. The
nontrivial result is the identification of this class, and the
demonstration that the Bessel-Hagen mechanism degenerates, for
Fierz-Pauli, into equivalence-class invariance rather than strict
local invariance.

The two broad ingredients of the result, namely the no-go for a
strictly gauge-invariant local $\partial h\,\partial h$ tensor and
the gauge-invariant equivalence class $[J^\mu_\zeta]$ that survives,
are not new in isolation: the obstruction to a strictly
gauge-invariant local spin-2 energy-momentum tensor is already
present in Deser and Henneaux~\cite{DeserHenneaux1995} and in the
related theorem of Magnano and
Sokolowski~\cite{MagnanoSokolowski2002}, while the quotient-space
interpretation is the standard characteristic-cohomology framework
of Barnich, Brandt and
Henneaux~\cite{BarnichBrandtHenneaux1995a,BarnichBrandtHenneaux1995b,BarnichBrandtHenneaux2000}.
What is new is that the present analysis brings these two
ingredients together in a single Bessel-Hagen treatment: the no-go
result explains why no strictly invariant local tensor can emerge,
while Eq.~\eqref{xi-shift} shows that the surviving Bessel-Hagen
object is a current class, with the particular field-equation
freedom associated with the compensating gauge parameter acquiring
a Noetherian origin in the boundary term $K^\mu_\xi$ of the inexact
gauge symmetry rather than the status of an external improvement
step.

Three nested classes are useful to distinguish here. (i) The
$\xi$-family $\{J^\mu_{\zeta,\xi}\}$ has members differing from
$J^\mu_{\zeta,0}$ only by terms proportional to the field
equations. (ii) Closing (i) under independent gauge
transformations $\delta_\chi$ of $h_{\mu\nu}$ adds the specific
superpotentials $\partial_\nu U^{[\mu\nu]}_{\zeta,\chi}$ of
Sec.~\ref{sec:class}. (iii) The standard BBH equivalence class
of $J^\mu_{\zeta,0}$ treats any antisymmetric superpotential and
any term proportional to the field equations as trivial. These
satisfy
$\text{(i)}\subset\text{(ii)}\subset\text{(iii)}$. The
$\xi$-family in (i), and its gauge orbit in (ii), should therefore
not be confused with the full improvement freedom. Named improved
expressions such as the Belinfante~\cite{Belinfante1940,Rosenfeld1940},
Hilbert~\cite{Hilbert1915}, Landau-Lifshitz~\cite{LandauLifshitz},
and M\o ller~\cite{Moller1961} tensors are best regarded, when related to the canonical tensor by
superpotentials and field-equation terms, as representatives of
the broader BBH class (iii). They need not be generated by varying
the compensating Bessel-Hagen parameter, nor by the specific
gauge-induced superpotentials of Sec.~\ref{sec:class}. The wider
improvement freedom of
Refs.~\cite{Baker2021,BakerKK2021,BakerLinnemannSmeenk2022,LinnemannSmeenkBaker2024}
therefore belongs to the full current-equivalence relation,
not to (i) or (ii).

The Bessel-Hagen construction is, in this sense, not a generator
of the full improvement family of spin-2 energy-momentum tensors;
its role is more diagnostic. It shows that the compensating gauge
freedom which is decisive in electromagnetism becomes, for the
Fierz-Pauli action, a weakly vanishing ambiguity. The remaining
differences among the familiar gravitational energy-momentum
representatives listed above arise from choices that lie outside
the Bessel-Hagen mechanism: relocalisation, Lagrangian
representative, background covariantisation, and choice of
pseudotensor superpotential. The Bessel-Hagen analysis
is orthogonal to these: it identifies the Noether current class
but does not, on its own, select a preferred local representative.

The gauge-fixing approach of Butcher, Hobson and
Lasenby~\cite{ButcherHobsonLasenby2010,ButcherLasenbyHobson2012,ButcherHobsonLasenby2012}
addresses the gauge ambiguity in a complementary way to the
Bessel-Hagen construction: rather than identify a gauge-invariant
equivalence class of currents, one selects a particular local
representative by imposing a gauge condition on $h_{\mu\nu}$. A
related first-derivative construction is T\'oth's Fierz-tensor
formulation of linearised gravity~\cite{Toth2022}, in which an
energy-momentum tensor with favourable properties, including
satisfaction of the dominant energy condition in a class of gauges
containing the transverse-traceless gauge, is obtained from the
Fierz tensor built from first derivatives of $h_{\mu\nu}$. This is
complementary to the present result: it belongs to the gauge-fixed
strategy rather than to the equivalence-class strategy of the
present paper.

The Bessel-Hagen analysis presented here is also the
linearised counterpart of the manifestly covariant Bessel-Hagen
construction of Hobson, Lasenby and
Barker~\cite{HobsonLasenbyBarker2024} for the gauge theories of Weyl
and of extended Weyl gravity, in which the local field strengths are
themselves gauge covariant and the construction succeeds in the
strict, rather than quotient, sense. A different generalisation of
the Bessel-Hagen method, to fourth-order conformally invariant
theories of gravity, has recently been given by
Faria~\cite{Faria2025}, who obtains a (gauge-dependent)
pseudo-tensor; the higher-derivative structure of that case differs
essentially from the second-order Fierz-Pauli setting considered
here. For broader context on the structure of massive and massless
spin-2 theory,
including gauge-invariance issues, see the review by Hinterbichler~\cite{Hinterbichler2012}.

\section{Conclusions}

In summary, the spin-2 Bessel-Hagen calculation produces no
strictly gauge-invariant local $\partial h\,\partial h$ tensor,
because $\partial h$ can be set to zero pointwise by a spin-2 gauge
transformation. Once the inexact nature of the Fierz-Pauli gauge
symmetry is treated correctly, however, the construction identifies
not a single tensor but an entire equivalence class of Noether
currents --- gauge invariant as a class, not pointwise. The
identity
$J^\mu_{\zeta,\xi}=J^\mu_{\zeta,0}-2{\cal E}^{\mu\nu}\xi_\nu$ of
Eq.~\eqref{xi-shift} describes the $\xi_\mu$-generated part of
this class: changing the compensating Bessel-Hagen gauge parameter
moves the current only by a weakly vanishing term. Gauge
transformations of the field further add superpotential
terms, and the full conserved-current class is the standard BBH
class of $J^\mu_{\zeta,0}$ modulo superpotentials and field
equations. The canonical Noether tensor, which at
linearised order coincides with the Einstein pseudotensor in the
usual $\Gamma\Gamma$ representation of the action, is a convenient
representative of this class. Other improved expressions in the
literature, when they differ by superpotentials and field-equation
terms, represent the same conservation law. The standard BBH
current-equivalence relation places these representatives on the
same footing, while the Bessel-Hagen construction identifies this
class from within a single inexact Noether calculation. This is precisely the
structure that the cohomological treatment of Barnich, Brandt and
Henneaux and the improvement-tensor literature had suggested must
be present, here derived directly at the Lagrangian level, with
the inexact gauge boundary term
$K^\mu_\xi$ supplying a Noetherian origin for the field-equation
freedom that is usually introduced by hand. The Bessel-Hagen method
therefore has a precise spin-2 counterpart for the Fierz-Pauli
action, but only as a gauge-invariant equivalence class of Noether
currents, not as a preferred local gauge-invariant
energy-momentum tensor.

In closing, we note that the Bessel-Hagen construction presented here
also extends to higher spins via the Fronsdal
action~\cite{Fronsdal1978}, the generalisation of (massless)
Fierz-Pauli to totally symmetric rank-$s$ tensor fields. The spin-3
case will be set out in a future paper, in
which a unique two-parameter compensating gauge choice reduces the
form variation to a generalised Christoffel symbol and the
inexact-gauge boundary term retains its Noetherian form, with the
coefficient $-2{\cal E}^{\mu\nu}\xi_\nu$ of \eqref{xi-shift} replaced
by $-3{\cal E}^{\mu\nu\rho} \xi_{\nu\rho}$. The structural conclusions
of the present paper are further expected to carry over to arbitrary integer
spin, but this is a topic for future research.

\begin{acknowledgments}
W.~B. is grateful for the support of Girton College, Cambridge, Marie
Sk\l{}odowska-Curie Actions, and the hospitality of the Helsinki
Institute of Physics. Co-funded by the European Union (Physics for Future -- Grant Agreement No. 101081515). Views and opinions expressed are however those of the author(s) only and do not necessarily reflect those of the European Union or European Research Executive Agency. Neither the European Union nor the granting authority can be held responsible for them.
\end{acknowledgments}

\end{document}